\documentstyle[11pt,twoside,colloqOHP2010,epsfig]{article}
\renewcommand{\footnoterule}{}    
                      
\let\oldtabular=\tabular
\def\tabular{\footnotesize\oldtabular}
\begin{document}
\title{Radial velocity follow-up of \textit{CoRoT} transiting exoplanets}
\author{A. Santerne$^{1,2}$, M. Endl$^3$, A. Hatzes$^4$, F. Bouchy$^{5, 2}$, C. Moutou$^1$, M. Deleuil$^1$ \& the \textit{CoRoT} radial velocity team}
\affil{$^1$ Laboratoire d'Astrophysique de Marseille, CNRS \& Universit\'e d'Aix-Marseille, Marseille, France [alexandre.santerne@oamp.fr]}
\affil{$^2$ Observatoire de Haute-Provence, CNRS \& Universit\'e d'Aix-Marseille, St.~Michel l'Observatoire, France}
\affil{$^3$ McDonald Observatory, The University of Texas, Austin, TX 78712,USA}
\affil{$^4$ Th\"uringer Landessternwarte, Sternwarte 5, Tautenburg 5, D-07778 Tautenburg, Germany}
\affil{$^5$ Institut d'Astrophysique de Paris, 98bis boulevard Arago, 75014 Paris, France} 
\begin{abstract}
We report on the results from the radial-velocity follow-up program  performed to establish the planetary nature and to  characterize the transiting candidates discovered by the space mission \textit{CoRoT}. We use the SOPHIE at OHP,  HARPS at ESO and the HIRES at Keck spectrographs to collect spectra and high-precision radial velocity (RV) measurements for several dozens different candidates from \textit{CoRoT}. We have measured the Rossiter-McLaughlin effect of several confirmed planets, especially CoRoT-1b which revealed that it is another highly inclined system. Such high-precision RV data are necessary for  the discovery of new transiting planets. Furthermore, several low mass planet candidates have emerged from our Keck and HARPS data.
\end{abstract}  
\section{Introduction}
Transiting exoplanets are unique targets for which we can measure the planetary radius by high accuracy photometry when the planet passes in front of its host star and its mass and orbital characteristics (eccentricity, alignment with the spin of the star) by Doppler measurements of the host star. Thus it is possible to compute the mean density of the planet and to  model its internal structure or to  explore the composition and characteristics of its atmosphere (albedo, temperature, and atmospheric composition) by photometry or spectroscopy observations during the transit or the eclipse.\\

\begin{figure}[h!]
\begin{center}
\begin{minipage}[b]{.55\textwidth}
\centering \epsfig{file=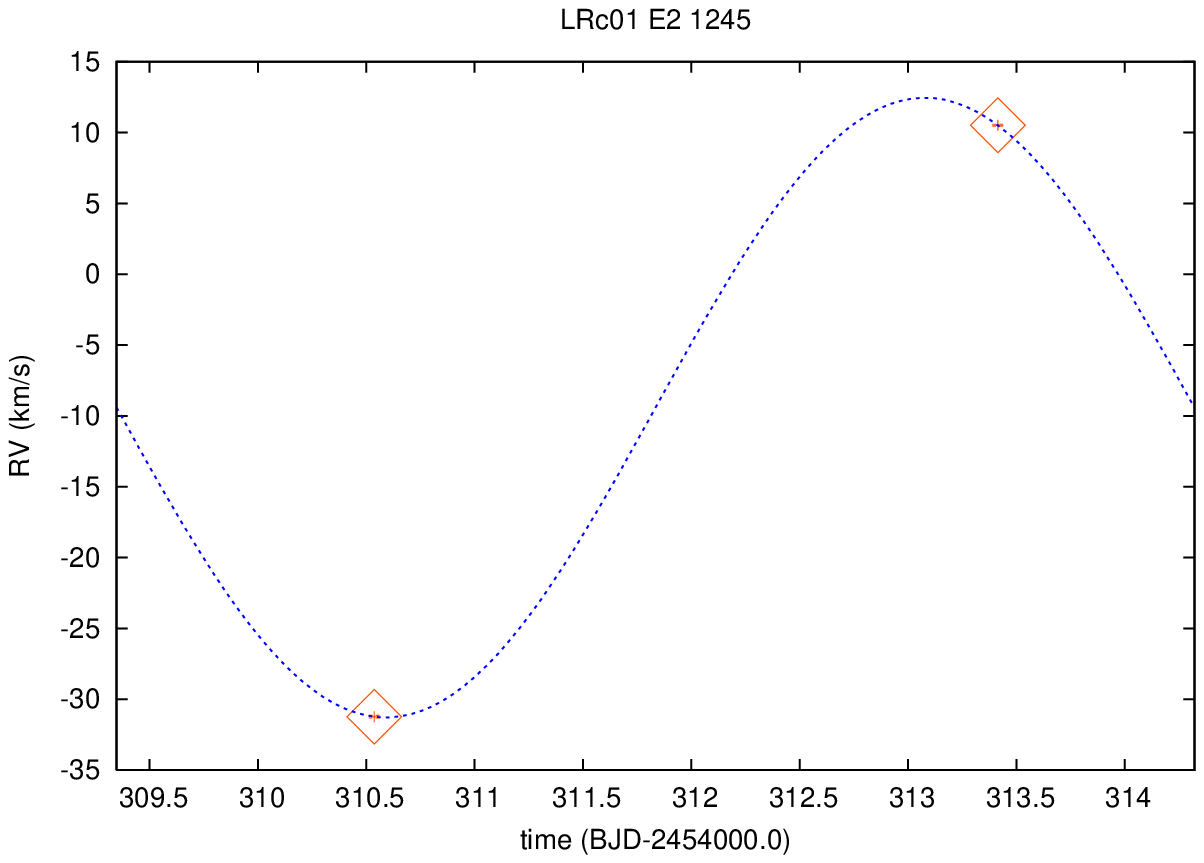, width=\textwidth} 
\end{minipage} \hfill
\begin{minipage}[b]{.3\textwidth}
\centering \epsfig{file=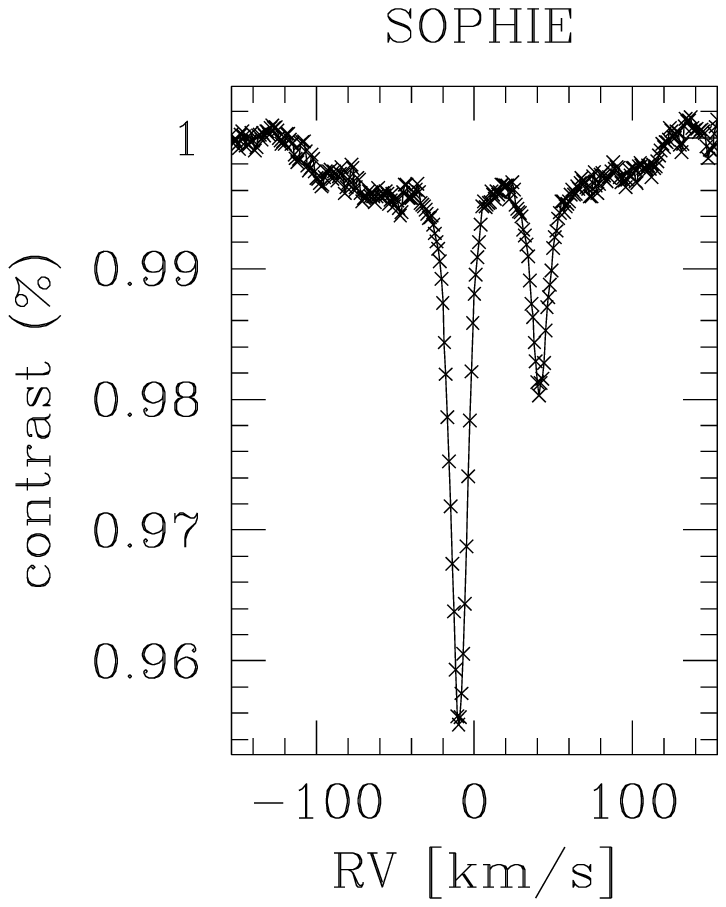, width=1\textwidth}
\end{minipage}
\label{transitscenarii}
\caption{(left) RV variations as function of time of the planetary candidate LRc01\_E2\_1245 (Cabrera et al. 2009) revealed as a SB1 by SOPHIE with only two measurements. RV are compatible with a companion of $m_c \sim 0.2 M_\odot$ with a period of 4.974 days, assuming $M_*=1~M_\odot$ and $i=90\deg$ and $e=0$.  (right) CCF$^*$ of a candidate revealed as a SB3 by SOPHIE. The third star have a large $v\sin i$}
\end{center}
\end{figure}

\textit{CoRoT} (Baglin et al. 2006; Deleuil et al., this book) is the first space-based photometric survey dedicated to finding transiting exoplanets. 
\textit{CoRoT} finds about 250 objects per run whose light curves show transit-like events. Most of them are clear eclipsing binaries (EB), but when a target shows periodic single transits (i.e. no secondary transits), no ellipsoidal variations and a shape, duration, and depth compatible with a transiting exoplanet, we consider it as a transiting exoplanet candidate. But these planetary candidates could still be mimicked by a transiting low-mass star, a grazing EB, a main sequence star eclipsing a giant star, or by an EB diluted by a third star. These EB scenarii (about 50\% of candidates) could be resolved by high-resolution spectroscopy observations in order to discard all binary scenarii (SB1, SB2, SB3, ...). For example, Fig. \ref{transitscenarii} shows the result of the CCF\footnote[1]{Cross-Correlation Function between the spectrum and a numeric mask used as template of a star spectrum. It corresponds to the mean spectral line.} of a transiting candidate followed-up with SOPHIE. When EB scenarii are discarded, precise RV observations are required to measure the mass of the transiting object and characterize its orbit.

\section{RV follow-up facilities}

Precise RV follow-up observations are obtained using a network of 3 spectrographs that share candidates in a optimized way depending on the brightness of the host star and the shallowness of the candidate:

\subsection{SOPHIE spectrograph}

SOPHIE (Bouchy et al 2009a) is a high-accuracy fiber-fed echelle spectrograph mounted on the 1.93-m telescope in Haute-Provence Observatory, France. SOPHIE has a resolution of $R_{HR}\sim 75,000$ (in High-Resolution mode) or $R_{HE}\sim 39,000$ (in High-Efficiency mode) at 550nm. Due to the faintness of the \textit{CoRoT} targets, we used only the High-Efficiency mode of SOPHIE which has an intrinsic stability of about $10\mathrm{m\,s^{-1}}$. We observe with SOPHIE all transiting planetary candidates up to V-magnitude of 15. About 13 nights per semester since 2007 are dedicated to the \textit{CoRoT} follow-up and permit to confirm and characterize some of the \textit{CoRoT} exoplanets discovered so far (e.g. Fig. \ref{HARPS-HIRES} on left).

\subsection{HARPS spectrograph}

HARPS (Mayor et al. 2003) is a high-accuracy fiber-fed echelle spectrograph mounted on the ESO 3.6-m telescope in La Silla Observatory, Chile. HARPS has a resolution of $R_{HAM}\sim 110,000$ (in High-Accuracy mode) or $R_{EGGS}\sim 70,000$ (in EGGS mode) at 550nm. We can follow targets up to V-magnitude of 16 but we focus on the shallowest candidates or on candidates with the longest period around the brightest targets. A large program of 16 nights per semester is dedicated to this program on HARPS and permitted to establish or confirm the planetary nature of most of the \textit{CoRoT} exoplanets discovered so far (e.g. Fig. \ref{HARPS-HIRES})

\subsection{HIRES spectrograph}

\begin{figure}[h!]
\begin{center}%
\begin{minipage}[b]{.45\textwidth} \centering \epsfig{file=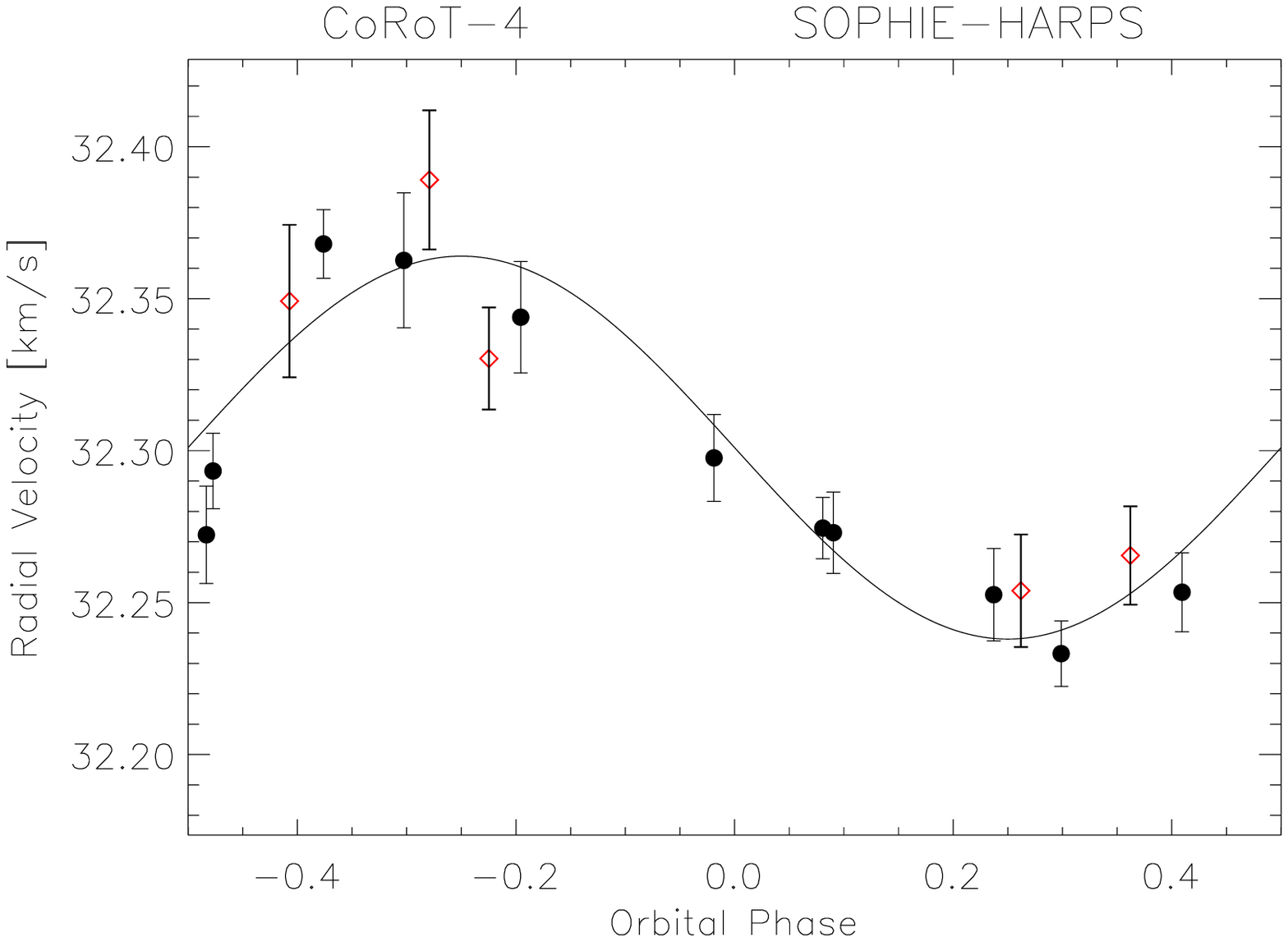, width=\textwidth} \end{minipage} \hfill
\begin{minipage}[b]{.45\textwidth} \centering \epsfig{file=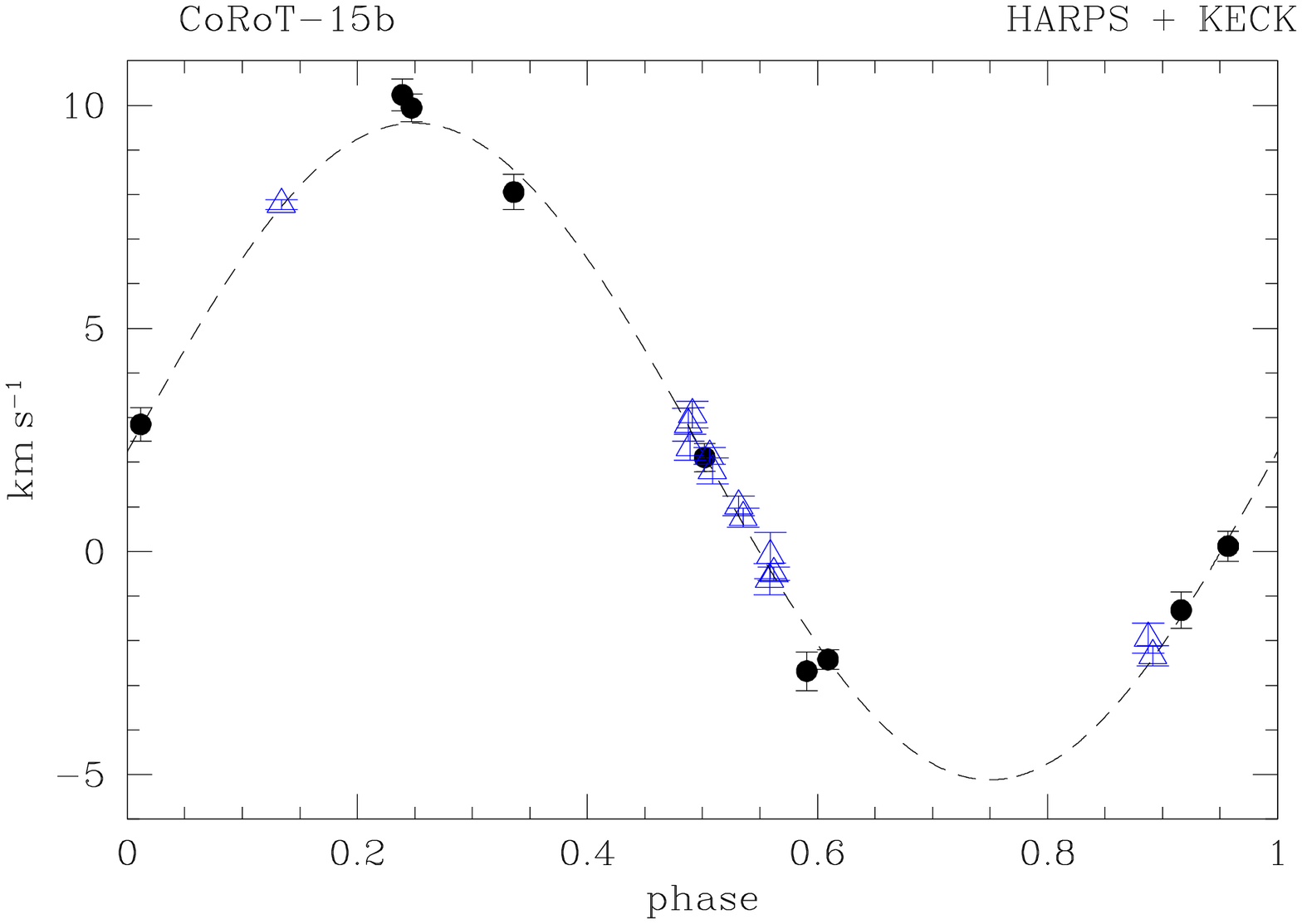, width=\textwidth} \end{minipage} \hfill
\caption{CoRoT-4b (left) (Moutou et al. 2008) and CoRoT-15 (right) (Bouchy et al. 2010) phase-folded RV curves characterized by both SOPHIE (red diamonds), HARPS (black circles) and HIRES (blue triangles).}
\label{HARPS-HIRES}
\end{center}
\end{figure}

The Keck/HIRES observations were obtained as part of NASA's key science project to support the \textit{CoRoT} mission. HIRES is a high-resolution, optical spectrograph (Vogt et al. 1994) mounted on the 10 m Keck 1 telescope on the summit of Mauna Kea, Hawaii. We use HIRES in combination with an iodine cell to obtain highly precise RV measurements. We use HIRES with the 0.86 arcsec slit that yields a resolving power of $\sim 45,000$ at 550nm. The 10-m aperture of Keck permits us to follow-up some candidates that are too faint for HARPS (V-magnitude $\sim 16$). In some cases, Keck/HIRES could observe Rossiter-McLaughlin (RM) anomaly with a better time resolution than HARPS or to obtain a high signal-to-noise (S/N) spectrum of the host star needed to characterize the star parameters. About 5 nights per semester are allocated on Keck for \textit{CoRoT} targets permit to confirm and characterize some of the \textit{CoRoT} exoplanets discovered so far (e.g. Fig. \ref{HARPS-HIRES} on right).

\subsection{Observations strategy}

Due to the large \textit{CoRoT} PSF\footnote[2]{Point Spread Function}, the first step of follow-up process is to re-observe the transit with a higher spatial resolution in order to discard all background eclipsing binaries (BEB) that could cause the transit (Deeg et al. 2009). Then, two high-resolution RV observations are scheduled at the extrema phase of the planetary orbit (assuming a circular orbit), corresponding to $T_0+P\left(n-\onequarter\right)$ and $T_0+P\left(n+\onequarter\right)$, where $P$ and $T_0$ are, respectively, the period and the epoch of the transit determined from the \textit{CoRoT} LC analysis and $n$ is the number of orbits since $T_0$. These two RV measurements are sufficient in most cases to estimate the nature of the transiting object: large RV variations of several $\mathrm{km\,s^{-1}}$ in phase with the \textit{CoRoT} ephemeris indicate an EB (SB1) as shown in Fig. \ref{transitscenarii}.  Small RV variations of less than a few $\mathrm{km\,s^{-1}}$ are compatible with a planetary nature and require more observations to be confirmed. 
\section{Limitations and RV diagnostics}
\subsection{Photon noise}

Mass characterization of low-mass planets strongly depends on the star brightness and its rotational velocity. Table \ref{photonoise} indicates photon noise uncertainties on SOPHIE and HARPS in one hour exposure time for 3 different $v\sin i$ and different V-magnitudes. Transiting Super-Earths ($K < 10 \mathrm{m.s^{-1}}$) can only be characterized around low-rotating stars brighter than mv=13, while transiting hot-Neptune  ($K < 30 \mathrm{m.s^{-1}}$) can be characterized with HARPS on stars up to mv=14. For stars fainter than mv=14, or fast-rotating stars, only giant planets and brown dwarves can be characterized.

\begin{table}[h]
\begin{minipage}[b]{\textwidth}\centering
\caption{SOPHIE and HARPS photon noise in 1 hour exposure time in $\mathrm{m\,s^{-1}}$}
\label{photonoise}
\begin{tabular}{|c|c|c|c|c|}
\hline
Spectrograph & mv & $v\sin i < 2 \mathrm{km\,s^{-1}}$ & $v\sin i \sim 5 \mathrm{km\,s^{-1}}$ & $v\sin i \sim 10 \mathrm{km\,s^{-1}}$\\
\hline
SOPHIE & 12 & 4 & 8 & 12\\
\tiny{HE mode} & 13 & 6 & 12 & 18\\
\tiny{$\sim 10 \mathrm{m\,s^{-1}}$} & 14 & 15 & 30 & 45\\
\tiny{systematics} & 15 & 30 & 60 & 90\\
 \hline
HARPS & 12 & 1.5 & 3 & 4.5\\
\tiny{HAM mode} & 13 & 2.5 & 5 & 8\\
 & 14 & 6 & 12 & 18\\
 & 15 & 15 & 30 & 45\\
 & 16 & 30 & 60 & 90\\
 \hline
\end{tabular}
\end{minipage}
\end{table}

\subsection{Contamination by Moon Background Light}

\begin{figure}[h!] 
\begin{center}%
\begin{minipage}[b]{.7\textwidth}
\centering \epsfig{file=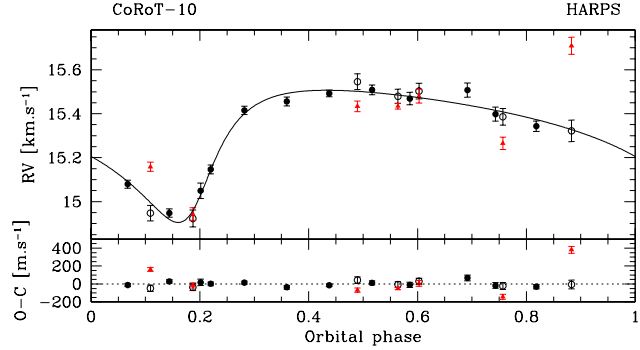, width=\textwidth} \end{minipage}
\caption{RV measurements from HARPS of the CoRoT-10 (Bonomo et al. 2010) host star. Full circles are observations not affected by the MBL, full triangles are observations affected by the MBL, open circles are observations corrected from the MBL.} \label{corot10}
\end{center}
\end{figure}

Due to the faintness of the \textit{CoRoT} target, observations are mostly scheduled in dark time (i.e., when the Moon is set). The Moon background light (MBL) that is blended the target spectrum can affect the RV measurement of up to several $\mathrm{km\,s^{-1}}$ as one can see in Fig. \ref{corot10}. The correction consists in subtracting the CCF of the sky observed simultaneously to the CCF of the target (for more details, see Bonomo et al 2010). 

\subsection{Blended eclipsing binary}

\begin{figure}[h!]
\begin{center}%
\begin{minipage}[b]{.64\textwidth}
\centering \epsfig{file=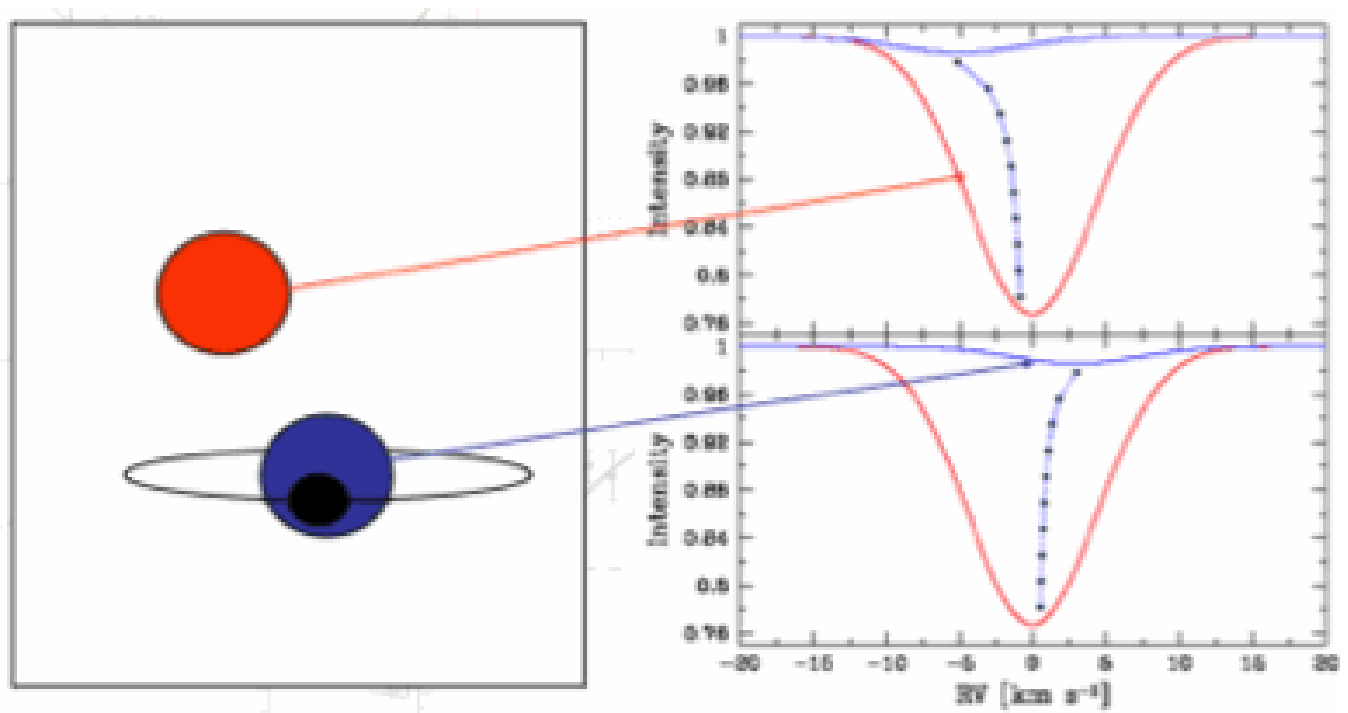, width=\textwidth} \end{minipage} \hfill
\begin{minipage}[b]{.34\textwidth}
\centering \epsfig{file=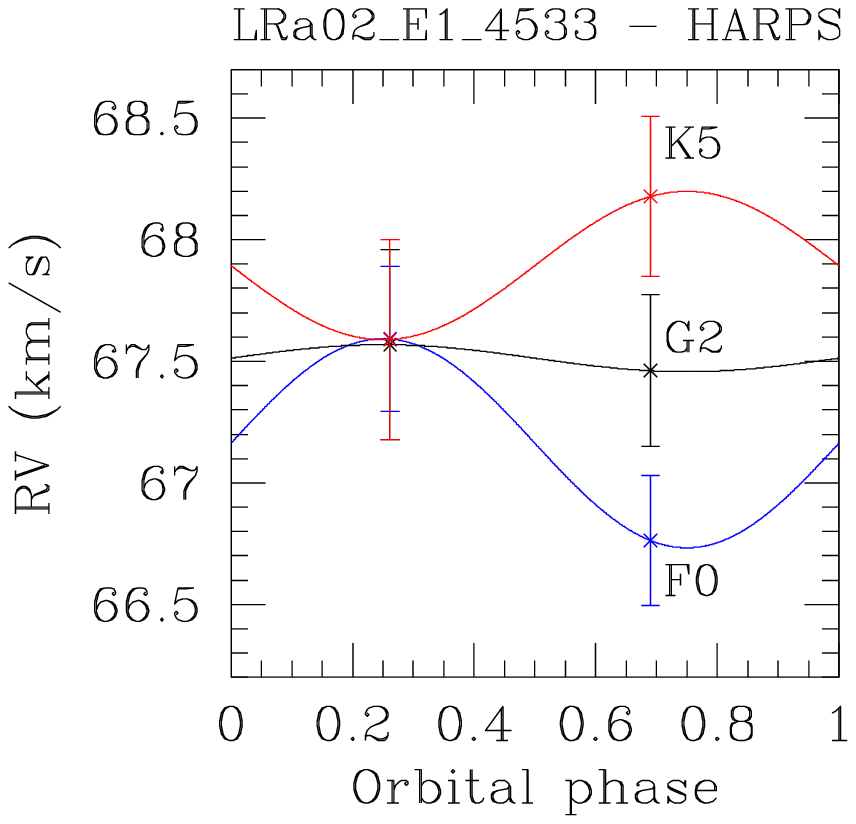, width=\textwidth}
\end{minipage}
\caption{(left) CCF asymmetry variations due to the presence of a large and narrow blended binary that introduced a correlation between bisector slope and radial velocity. (right) Background eclipsing binary revealed by mask effect between the RV obtained with the F0, G2 and K5 templates.} \label{4533} \label{BEB}
\end{center}
\end{figure}

In the case of a triple system or an unresolved BEB, the spectrum of the main star could be blended with the spectrum of the eclipsing binary and it can affect the RV measurements of the main star (Bouchy et al 2009b). This kind of scenario can mimic the RV signature of a planetary companion,  but produces asymmetries in the CCF in phase with the orbital period (see Fig. \ref{BEB}), or different RV amplitudes when computing the CCF with masks from different spectral types which can detect blending stars with different  $T_{eff}$ (see Fig. \ref{4533}). 

\subsection{Double binaries}

\begin{figure}[h!]
\begin{center}
\begin{minipage}[b]{.43\textwidth}\centering \epsfig{file=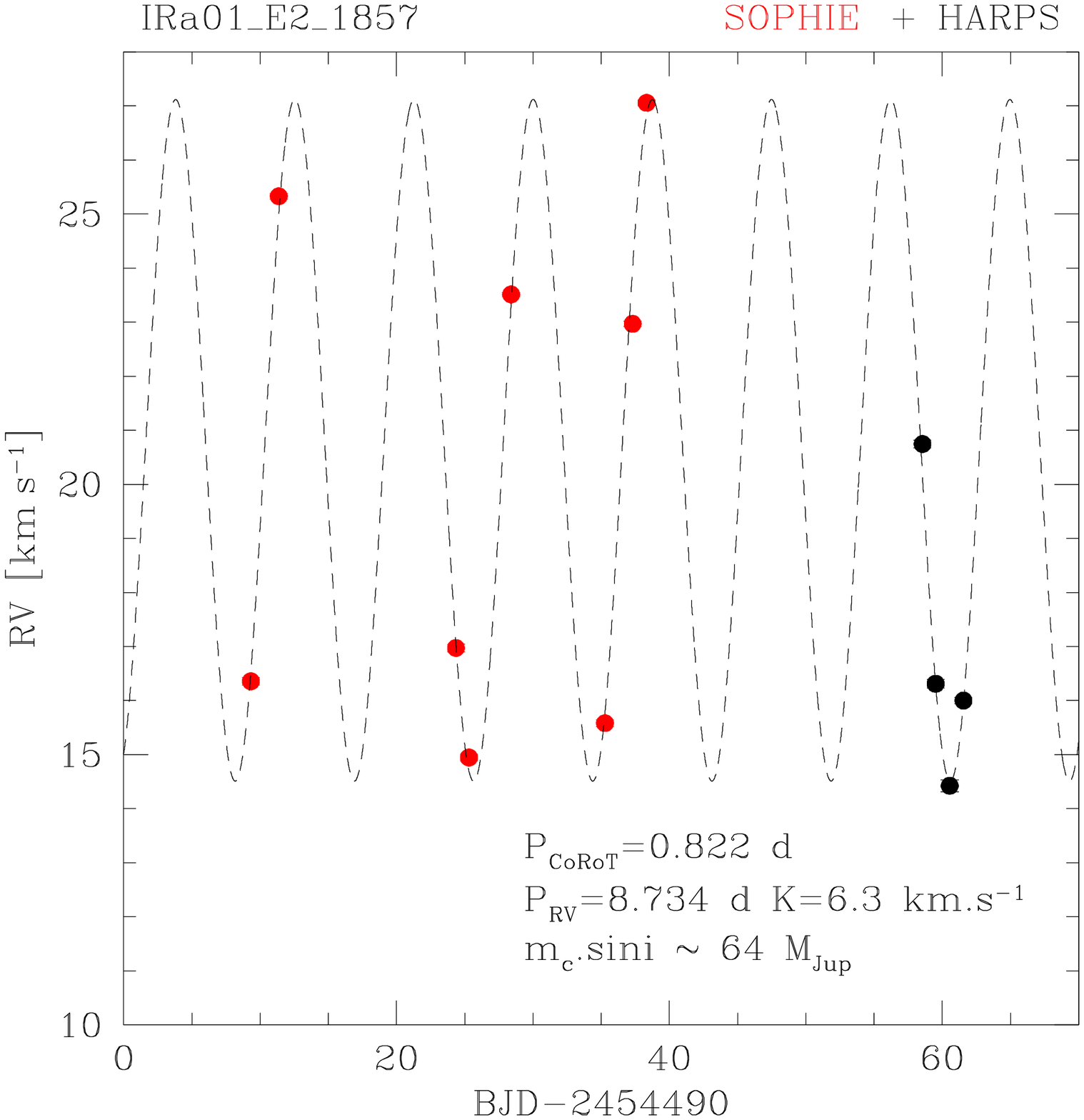, width=\textwidth}\end{minipage} \hfill
\begin{minipage}[b]{.4\textwidth}\centering \epsfig{file=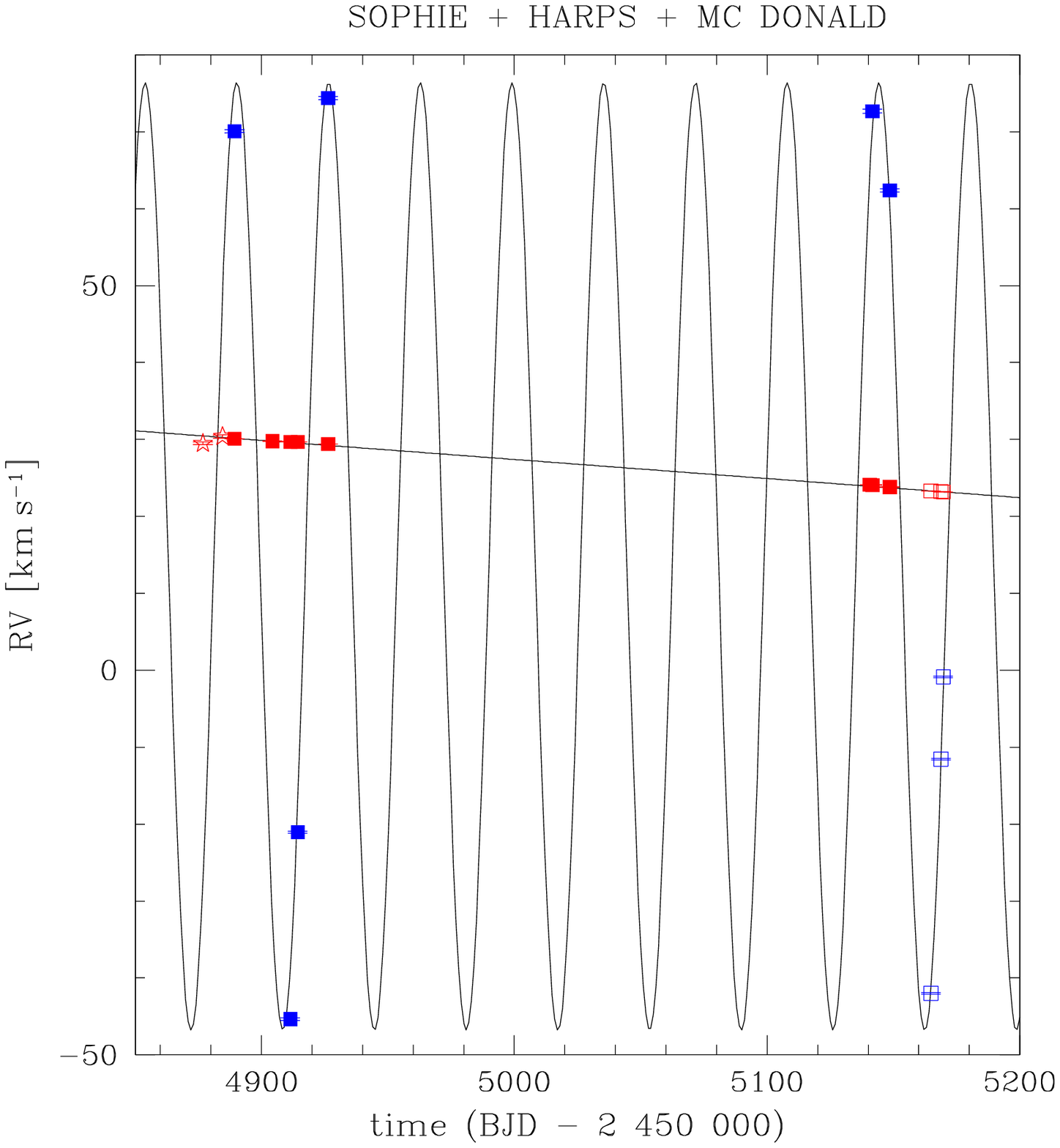, width=\textwidth}\end{minipage}
\caption{Two cases of double blended binaries solved by SOPHIE and HARPS: (left) brown dwarf at 8.73 days blended inside the \textit{CoRoT} photometric mask with a 0.8 days EB (Moutou et al. 2009). (right) long-period foreground binary blended inside the seeing with a BEB at 36.6 days. The two plots show RV variations as function of time overplotted with best RV-fit of the 8.73-days binary (left panel) and both the best RV-fit of the primary binary and the secondary binary (right panel).}
\label{murphy}
\end{center}
\end{figure}

In some cases, RV variations are not in phase with the transit seen by \textit{CoRoT}. This is mostly caused by blended binaries : a contaminating EB at the \textit{CoRoT} period and a binary with a different period. Fig.~\ref{murphy} shows two of these cases where the transit is explained by a BEB. For the first one (left panel), the background binary is separated by a few arcsec (within the \textit{CoRoT} photometric mask) and for the second, the two binaries are located inside the seeing  disk (Tal-Or et al., in prep).

\subsection{Stellar activity : the case of CoRoT-7}

\begin{figure}[h!]
\begin{center}
\begin{minipage}[b]{.8\textwidth}
\centering \epsfig{file=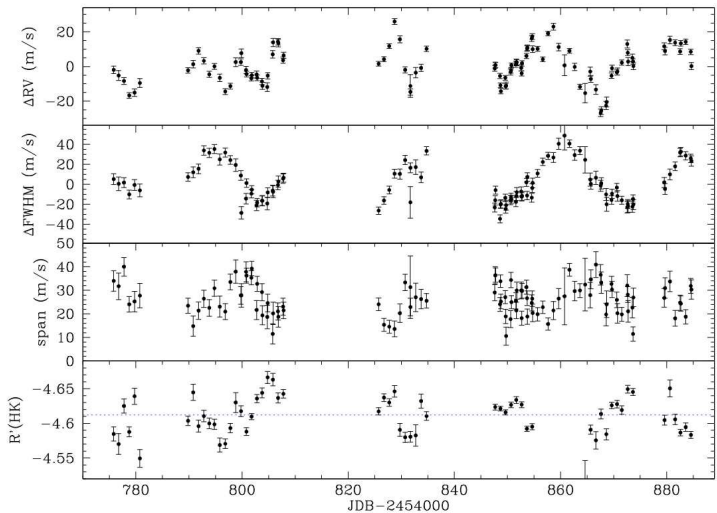, width=\textwidth}
\caption{All HARPS observations of the CoRoT-7 host star (Queloz et al. 2009) showing (from top to bottom) RV, FWHM, bisector span and CaII activity level (R'(HK))  measurements as a function of time. One can see that FWHM, bisector span and CaII activity level present significant variations that could be explained by stellar activity which affects also RV measurements.}
\label{corot7}
\end{minipage}
\end{center}
\end{figure}

CoRoT-7 is an active star which hosts at least 2 super-Earths. Stellar activity induces RV signatures (see Fig. \ref{corot7}) at the level of a few $\mathrm{m.s^{-1}}$. To dissociate stellar activity from planetary signature, it is necessary to have ancillary measurements like simultaneous photometry, bisector, FWHM or CaII measurements. Right now, characteristics of the CoRoT-7 system is still under intense discussion (see Queloz et al. 2009; Hatzes et al. 2010; Lanza et al. 2010; Pont et al. 2010b; Boisse et al. subm., Ferraz-Mello et al, subm.).

\section{Follow-up results}
\subsection{About follow-up statistics}

\textit{CoRoT} so far has discovered 15 new transiting exoplanets and brown dwarves (Deleuil et al. this book). These discoveries are the fruit of \textit{CoRoT} high-accuracy space-based photometry combined with an intensive ground-based photometric and spectroscopic follow-up of about 200 transiting exoplanet candidates for the first 3 years of \textit{CoRoT}. Run report papers (IRa01: Moutou et al. 2009, LRc01: Cabrera et al 2009, SRc01: Erikson et al in prep., LRa01: Carone et al. in prep., LRc02: Bord\'e et al. in prep.) present  the observation and analysis sub-sampled of these candidates including about 50\% of binary (SB1, SB2, BEB, blended EB), about 10\% of hot stars for which we cannot measure RV with enough precision to characterize the candidate's mass (and could still be planetary), and about 10\% of confirmed transiting exoplanets. The remaining 30\% are the faintest stars or too low-priority candidates with poor planet likelihood which were not followed up or unsolved candidates due to photon noise limitation.

\subsection{Rossiter-McLaughlin (RM) observations}

To complete the characterization of the system, we observe Rossiter-McLaughlin RV anomaly (Winn et al., this book; Triaud et al., this book) during the transit. It permits us to measure the sky-projected angle between the spin of the star and the orbit of the planet. CoRoT-2b (Bouchy et al. 2008) and CoRoT-3b (Triaud et al. 2009) are spin-orbit aligned exoplanets while CoRoT-1b was revealed to be a misaligned planet (Pont et al. 2010). Observation of part of the transit of CoRoT-11b indicates another misaligned exoplanet (Gandolfi et al. 2010), although observations did not cover the complete transit. Part of the transits of CoRoT-9b and CoRoT-6b were observed by SOPHIE and HARPS and require more observations to conclude.

\begin{figure}[h!]
\begin{center}%
\begin{minipage}[b]{.65\textwidth} \centering \epsfig{file=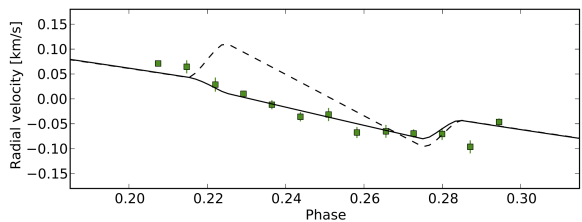, width=\textwidth} \end{minipage} \hfill
\caption{RM observations of CoRoT-1b by HIRES (Pont et al. 2010a) with the best-fit in solid line for $\lambda =77\deg$. The dashed line shows what we expected for an aligned spin-orbit ($\lambda = 0\deg$).}
\label{RM1}
\end{center}
\end{figure}

\section{Conclusion}

Transiting exoplanet surveys need RV follow-up in order to determine the nature and the characteristics of the exoplanets candidates. Using the facilities of an optimized network of 3 high-resolution spectrographs for follow-up (SOPHIE, HARPS and HIRES) with powerful diagnostics to discard false positives and secure detection, \textit{CoRoT} is, so far, the photometric survey that has discovered more planets per square degree of observed sky. More than 1000 spectra with signal-to-noise  of up to 100 on about 200 transit candidates were taken with SOPHIE, HARPS, and HIRES during the first 3 years of \textit{CoRoT}. Fifteen new exoplanets and brown dwarves have been discovered and characterized by these high-resolution spectrographs so far. Currently, 6 of these \textit{CoRoT} planets  have been observed in order to measure their RM effect. One planet orbit is is clearly misaligned with the spin of its host star while  another one shows strong  evidence of a misalignment, but requires more measurements to confirm this.

\acknowledgments{The data presented herein were obtained at the OHP, ESO and at W.M. Keck Observatory from telescope time allocated to the National Aeronautics and Space Administration through the agency's scientific partnership with the California Institute of Technology and the University of California. The Observatory was made possible by the generous financial support of the W.M. Keck Foundation. The full RV team is made of William D. Cochran, Rodrigo F. D\'iaz,  Davide Gandolfi, Eike Guenther, Guillaume H\'ebrard, Christophe Lovis, Phillip J. MacQueen, and Didier Queloz.
}

\end{document}